\documentclass[showpacs,prb,preprintnumbers,amsmath,amssymb,twocolumn]{revtex4-2}

\usepackage{graphicx}
\usepackage{dcolumn}
\usepackage{bm}
\usepackage{hyperref}

\begin{document}
\title{Evidence for magnetic clusters in stoichiometric quantum critical CeRu$_2$Si$_2$}

\author{Alex Breta\~{n}a, Sean Fayfar,  and Wouter Montfrooij} 
\affiliation{ Department of Physics and Astronomy, University of Missouri, Columbia, Missouri 65211, USA}
\begin{abstract}
{Systems that have been prepared to undergo a second-order phase transition at zero Kelvin, the so-called quantum critical systems, appear to fall into two categories: (chemically) heavily-doped systems where the unusual properties can be related to a disorder-induced distribution of Kondo shielding temperatures, and (almost) stoichiometric systems where the departures from Fermi-liquid theory have been attributed to intrinsic instabilities. Here we show that this distinction is not as clear cut and that magnetic clusters associated with a distribution of Kondo shielding temperatures are also present in CeRu$_2$Si$_2$, a system close to a quantum critical point. By revisiting published data on this system and comparing them to the results for heavily-doped quantum critical Ce(Ru$_{0.755}$Fe$_{0.245}$)$_2$Ge$_2$, we show that clusters exist in both systems at low temperatures, and that the moments of the Ce-ions within these clusters have all lined up with their neighbors. This implies that the dominant physics that drives heavily-doped systems, namely spontaneous formation of magnetic clusters, should also play a leading role in the response of homogenous systems. This represents a notable departure of how the physics that governs quantum critical points is treated in the literature.}
\end{abstract}
 \maketitle
\section{Introduction}
Metallic systems that house magnetic ions in their unit cells are known\cite{stewart} as Kondo lattice systems. In contrast to metals diluted with a few magnetic impurities, the physics behind the low temperature behavior of the Kondo lattices remains unresolved. Of particular interest are the Kondo lattice systems that display a competition between the Kondo shielding tendencies of the magnetic moments by the conduction electrons and long-range order of the magnetic moments through the conduction mediated  Ruderman$-$Kittel$-$Kasuya$-$Yosida (RKKY) interaction\cite{rkky,rkky2,rkky3}. When these systems are tuned in such a way that this competition remains unresolved to zero Kelvin, then it is experimentally observed that the low-temperature response does not follow the predictions\cite{philip} of Landau Fermi liquid theory. Instead, non-Fermi liquid behavior manifests itself in the specific heat, uniform susceptibility, resistivity, as well as in the dynamic susceptibility where $E/T$-scaling has been observed\cite{meigan,schroder,meigan2,wouterprl} in some systems. Currently, we do not have an overarching theory to qualitatively explain the many observations on these quantum critical systems, that is, systems that have been prepared to be on the verge of ordering at zero Kelvin, the so-called quantum critical point (QCP).\\

The two competing tendencies in Kondo lattices are well understood at the atomic level, however. The ability of conduction electrons to form a magnetic singlet with the moment of a magnetic ion was first described by Kondo\cite{kondo} and treated using renormalization theory by Wilson\cite{wilson}. Similarly, Ruderman$-$Kittel$-$Kasuya$-$Yosida  showed\cite{rkky,rkky2,rkky3} how magnetic moment alignment can be mediated from ion to ion by the conduction electrons. Both mechanisms originate from the same physical exchange interaction between the electrons of the local atomic orbitals that give rise to a magnetic moment and the extended conduction electron orbitals. The strength of this exchange interaction $J$  depends highly sensitively on the degree of overlap between these orbitals. If the overlap is large, then the ground state of the metal will be one where the magnetic moments have been shielded by the conduction electrons, and no long-range order is observed. For weaker overlap, the moments can persist down to zero Kelvin, and long-range order ensues at a finite temperature. When the two tendencies compete with each other down to zero Kelvin, then the system will display critical behavior upon approaching the order-disorder transition. This critical behavior is observed both as a function of temperature as well as external parameters such as hydrostatic pressure and magnetic fields. Since phase transitions at zero Kelvin are not driven by thermal fluctuations, they should fall into new universality classes.\cite{sachdev} Such second order phase transitions are referred to as quantum phase transitions.\\

In practice, systems that have been tuned to be at the QCP fall into two categories, and our efforts at theoretical understanding of the low temperature response of these systems reflect this dichotomy. On the one hand there exist (nearly) stoichiometric compounds whose composition is already so close to a QCP that they can be fine-tuned to be exactly at the quantum critical point by applying hydrostatic pressure, or by a small amount of chemical substitution to expand or shrink the lattice, thereby affecting the degree of orbital overlap in order to achieve the fine-tuning. Naturally, these systems have received the bulk of theoretical attention and various theories have been forwarded to describe the low-temperature response either as a localized instability against moment formation,\cite{si} or as a collective instability againt ordering rooted in the topology of the Fermi surface.\cite{hertz,millis} On the other hand,  systems far from a QCP can still be tuned,\cite{stewart} but now chemical pressure is required, resulting in high levels of chemical dopants, introducing a high degree of disorder. Given the high degree of disorder, theoretical efforts have focused on disorder related effects, such as a distribution of Kondo shielding temperatures and manifestations of a Griffith phase\cite{neto} where rare ordered subvolumes of the sample have a disproportionate influence on the overall response. A perusal of the literature on quantum critical systems leads one to believe that heavily-doped systems do not shed light on the true nature of quantum phase transitions in non-disordered stoichiometric systems. This may be partly due to the success\cite{bernal,minerva} of the Kondo disorder model, as detailed below. However, in this paper we show that effects normally associated with chemical disorder play a large, perhaps even dominating role in stoichiometric systems as well.\\

In the Kondo disorder model,\cite{bernal,minerva} the response of a heavily-doped quantum critical system is described as the response of a collection of non-interacting magnetic ions that are being Kondo shielded at different temperatures. The advantage of this model is that, given a distribution of Kondo shielding temperatures $P(T_K)$, the known expressions\cite{tabata} for susceptibility and specific heat for dilute magnetic moments can be applied and weighed with the overall distribution in order to arrive at the response of the doped system at any temperature. The parameters of the distribution can then be fine-tuned (fitted) to arrive at the best overall agreement with the experimental results for susceptibility and specific heat. Moreover, the idea of there being a distribution of Kondo temperatures in the first place has a solid physical foundation as the Kondo temperature depends exponentially on the degree of overlap of neighboring orbitals, which in turn depends on the interatomic separation\cite{endstra} $r$ as $\sim$ 1/$r^{10}$ or $\sim$ 1/$r^{12}$. This model was successfully applied\cite{bernal} to quantum critical UCu$_4$Pd and UCu$_{3.5}$Pd$_{1.5}$ where the susceptibility and specific heat were modeled simultaneously by one Gaussian distribution of the exchange interaction $J$, representing the physical overlap between localized moment-carrying orbitals and extended conduction electron orbitals. Applications to other disordered systems also showed that disorder plays a role, although it was not always possible to model the susceptibility and specific heat with one set of parameters such as when modeling\cite{tabata} heavily-doped CeRuRhSi$_2$.\\

When describing stoichiometric systems that are (very) close to a quantum critical point, such as CeRu$_2$Si$_2$, CeCu$_6$, YbRh$_2$Si$_2$, disorder is not taken into account. Instead, the experimental observations are compared to the theoretical predictions, such as the spin density wave scenario,\cite{millis,moriya} local moment scenario,\cite{schroder,si} or the self-consistent spin renormalization (SCR) model.\cite{moriya} We will not detail the various scenarios here; none of the scenarios describe all quantum critical systems, but each scenario appears to capture the essence of particular systems quite well. As mentioned, these theories tend to be applied solely to (near) stoichiometric systems while the response of heavily-doped systems are viewed as being disorder driven.\\

In this paper, we argue that ignoring the role of disorder in stoichiometric systems from the outset is not justified for two reasons. First, stoichiometric systems do possess a distribution of Kondo shielding temperatures, albeit a dynamic one. The amplitude of the zero point motion ($\sim$ 0.05 {\AA}) is comparable to doping induced changes\cite{fontes} in atomic separations, and as a result, a stoichiometric system will not have a uniform Kondo temperature either. Second, if the low temperature response of disordered quantum critical UCu$_4$Pd can indeed be described as being the result of the individual response of non-interacting magnetic moments being shielded by the conduction electrons, then this would imply that coherent collective effects do not play a significant role in these systems. As such, stoichiometric quantum critical systems should not display any interesting behavior either, by and large, else we would have found that disorder alone is not sufficient to describe the quantum critical physics in heavily-doped systems. Thus, we either have to accept that it is fortuitous that the Kondo disorder model describes heavily-doped systems as well as it does, or we have to accept that the response of a system near a quantum phase transition is inherently and fundamentally different between a stoichiometric and a doped system. In here, we show that this is not the case: we show that the effects of a static distribution of Kondo shielding temperatures induced by chemical doping are very similar to the effects of a dynamic distribution of Kondo temperatures originating from zero point motion. We use the (near) stoichiometric system CeRu$_2$Si$_2$ and heavily-doped quantum critical Ce(Ru$_{0.755}$Fe$_{0.245}$)$_2$Ge$_2$ to make our case.\\ 

We have chosen stoichiometric CeRu$_2$Si$_2$ as our reference system as this compound has been investigated by many groups.\cite{flouquet,tabata,takahashi,kadowaki,kadowaki2} In particular, we use the data gathered by Tabata\cite{tabata} on susceptibility and specific heat measurements, and the neutron scattering data on single crystal samples by Kadowaki {\it et al.}\cite{kadowaki,kadowaki2}\\

The organization of this paper is as follows. In the section on theory we review how zero point motion induces a distribution of Kondo shielding temperatures, and we introduce a model based on experimental observations that indicates that stoichiometric systems indeed have a significant distribution of Kondo temperatures. We also review the tell-tale signs of disorder playing a role in the specific heat, in the uniform susceptibility, and in neutron scattering experiments.  In the results section, we make a direct comparison between quantum critical Ce(Ru$_{0.755}$Fe$_{0.245}$)$_2$Ge$_2$ and stoichiometric CeRu$_2$Si$_2$.  In the discussion section, we scrutinize the likelihood that disorder actually plays a role in the quantum critical response of CeRu$_2$Si$_2$, and we discuss an alternative interpretation of the Wilson ratio.

\section{Theory}
\subsection{Distribution of Kondo temperatures}
The Kondo temperature\cite{kondo} $T_K$ depends\cite{philip} on the bandwidth $D$ of the conduction band, the density of states $\rho$ at the Fermi level, and the exchange coupling $J$; it is associated with the temperature below which it becomes energetically favorable for the conduction electron(s) to form a singlet with the magnetic impurity and is approximated as\cite{philip}
\begin{equation}
T_K =De^{-\rho|J|}
\label{tk}
\end{equation}
Wilson showed\cite{wilson} that Kondo shielding is not complete until $T$ = 0 K, however, the temperature scale of Eq. \ref{tk} still serves as an indication of when moments become shielded enough to no longer be able to align with neighboring moments. In their description of the non-Fermi liquid behavior of UCu$_4$Pd andUCu$_{3.5}$Pd$_{1.5}$, Bernal {\it et al.}\cite{bernal} used a Gaussian distribution for $\lambda = \rho |J| $ to model their data. Thus, they used three free parameters ($D$, $\langle\lambda\rangle$, and $w$, the width of the Gaussian) to model their susceptibility and specific heat data for each composition. In here, we introduce a slight improvement on this distribution by using one that is based on the deviations of ions from their equilibrium lattice positions. This deviation can be directly measured using neutron scattering experiments and goes by the name of Debye-Waller factor.\\

In order to relate changes in inter-ionic separations to the exchange interaction $J$, we follow the procedure outlined in Endstra {\it et al.}\cite{endstra} where it was shown that for a given pair of ions the exchange interaction depends on their separation $r$  as 
\begin{equation}
J \sim V_{df}^2/(E_{F}-E_f) \sim 1/r^{12},
\label{v}
\end{equation}
where $V_{df}$ is the hybridization matrix element between an f-orbital and a d-orbital, $E_F$ is the Fermi energy and $E_f$ the the energy level of the f-electron. Endstra {\it et al.} assumed\cite{endstra} that the variation in $E_{F}-E_f$ could be neglected, resulting in the 1/$r^{12}$ dependence of $J$ on inter-ionic separation.\\

For our distribution of Kondo temperatures, we use a Gaussian distribution of inter-ionic separations $r$. Such Gaussian distributions are borne out by experiments\cite{squires} and represent approximating the motion of ions around their equilibrium positions $r_{eq}$ as a harmonic motion. We arrive at the following distribution for Kondo temperatures $P(T_K)$ based on our distribution of separations $P(r)$ using $A$ to capture  the constant of proportionality in Eq. \ref{v} as well as $\rho$, and $r_c$ the average center-to-center distance between the two ions:
\begin{equation}
P(r) =\frac{\sqrt{ln2}}{\sqrt{\pi}\sigma}e^{-(r-r_{c})^2/\sigma^2 ln2};\quad 
T_K =De^{-r^{12}/A}.
\end{equation}
From this we can relate the two distributions as
\begin{equation}
P(T_K) =P(r) |\frac{dr}{dT_K}| =P(r)\frac{A}{12r^{11}T_K}.
\label{ptk}
\end{equation}
The advantage of this distribution over the one employed by Bernal {\it et al.}\cite{bernal}  is that it does not come with an artificial ln($T_K$) divergence which is the result of assuming a Gaussian distribution for $\lambda$. However, this assumption does not properly reflect the dependence of $T_K$ on inter-ionic separation as Eq. \ref{ptk} does. It is possible to arrive at a reasonable estimate for how the parameters in our distribution ($A$, $\sigma$) relate to those of the Bernal distribution ($<\lambda>$, $w$) by equating the mean Kondo temperature and the width of the distribution, yielding:
\begin{equation}
A =\langle\lambda\rangle r_c^{12};\quad \frac{\sigma}{r_c} =\frac{w}{12\langle\lambda\rangle }.
\end{equation}
For example, using the parameters published for UCu$_4$Pd\cite{bernal} and the measured\cite{ucu} U-Cu distance of 3.49 \AA, we find $\sigma$ = 0.049 \AA. Using the parameters listed by Tabata\cite{tabata} for CeRuRhSi$_2$, we find $\sigma$ = 0.038 \AA. These are reasonable values for the Debye-Waller factor at low temperatures: for instance, the measured\cite{bouchette} isotropic Debye-Waller factors for Ce and Ru in CeRu$_2$Si$_2$ are 0.039 {\AA}  and 0.058 {\AA}, respectively (although reported with large uncertainties). Note that since the Debye-Waller factor is experimentally accessible, this reduces the number of adjustable parameters in the distribution of Kondo shielding temperatures.\\

\begin{figure}[tbh]
    \begin{center}
    \includegraphics*[viewport =160 125 580 520,width =85mm,clip]{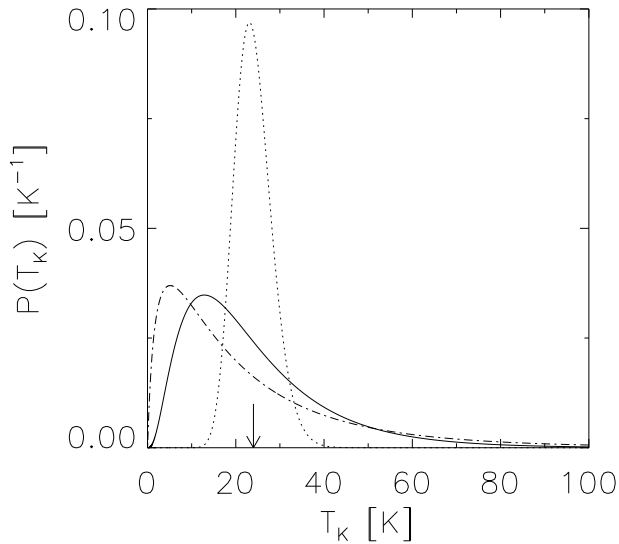}
    \end{center}
    \caption{The distribution of Kondo temperatures according to Eq. \ref{ptk} for three values of $\sigma$: $\sigma$ = 0.01 \AA (dotted curve), $\sigma$ = 0.039 \AA (solid curve), $\sigma$ = 0.058 \AA (dash-dotted curve). All distrubutions have the same average Kondo temperature of 24 K (arrow).} 
    \label{distribution}
\end{figure}

In Fig. \ref{distribution} we show the possible distributions of Kondo shielding temperatures based on the measured Debye-Waller factor, the reported value for $\langle\lambda\rangle$ = 0.18 for CeRuRhSi$_2$, and the average Kondo $\langle T_K\rangle$ = 24 K for CeRu$_2$Si$_2$. Whereas this value for  $\langle\lambda\rangle$ will not be exactly correct for stoichiometric CeRu$_2$Si$_2$, the figure offers a good indication of what an instantaneous distribution of Kondo temperatures looks like in a disorder-free system. As can be seen in this figure, for realistic values for the amplitude of zero point motion of the order of $\sim$ 0.05 {\AA}, the instantaneous distribution of Kondo temperatures is very wide. It is only for unphysical values of $\sigma$ $\sim$  0.01 {\AA}, as opposed to the two realistic values shown in the figure, that the distribution is sharp at $T$ = 0 K. What is clear from this figure is that theoretical approaches describing quantum critical physics in stoichiometric systems wherein the Kondo temperature is treated as a uniform parameter throughout the lattice and throughout time are simply not justified as  a starting assumption.\\

For heavily-doped systems, the distribution in atomic separations is, at least partially, static. Randomly substituting smaller or larger ions locally leads to lattice shrinking and expansion, necessarily leading to a permanent distribution of inter-ionic separations. Of course, the time-dependent changes are also present in heavily-doped systems. For simplicity, in this paper we treat the distribution of Kondo temperatures in heavily-doped systems as if they were entirely static, an assumption based on what was observed in neutron scattering experiments.\cite{fridge}\\ 

Whether the instantaneous distribution of Kondo temperatures is actually reflected in experimentally accessible quantities is a matter of timescales. Should the electronic timescale associated with Kondo shielding of moments be fast enough that it appears as if this ever-changing distribution were static, then we could see the effects of the distribution in our experiments. On the other hand, if the electronic timescales are slower than those associated with zero point motion, or if (the timescales of) our experiments are such that our probes only measure time-averaged values, then we would not see a reflection of this instantaneous disorder. We will address this issue in the next subsection.

\subsection{Experimental signature of disorder}
Should Kondo disorder play a role in the low-temperature response of stoichiometric CeRu$_2$Si$_2$, then neutron scattering should be able to directly observe this. The hallmark consequence of a distribution of Kondo temperatures is the appearance of magnetic clusters whose moments have all lined up with their neighbors. This is detailed in reference [\onlinecite{fridge}], and in a nutshell the reasoning is the following. Upon lowering the temperature, local moments are shielded at random, resulting in a fragmentation of the magnetic lattice. When clusters of moments peel of from the main, lattice spanning collection of moments, then the moments on these clusters will align\cite{heitmannconf} with their neighbors because of finite-size effects: an order destroying fluctuation on the cluster simply costs too much energy as the longest wavelength possible for such a fluctuation is limited by the size of the cluster, and the energy cost of a fluctuation is inversely proportional to its wavelength. As a result, clusters that become isolated will order and show up in scattering experiments. This behavior was used to identify clusters in the first place in heavily-doped quantum critical Ce(Ru$_{0.755}$Fe$_{0.245}$)$_2$Ge$_2$.\\

Both CeRu$_2$Si$_2$ and Ce(Ru$_{0.755}$Fe$_{0.245}$)$_2$Ge$_2$ are 3d-Ising systems to very good approximation,\cite{isingnature} and both systems have an identical (122) tetragonal structure.\cite{122} Since clusters result from random shielding of moments, the experimental signature of ordered clusters are short-range magnetic correlations that span equal number of moments along different crystallographic directions. In contrast, standard short-range order associated with incipient magnetic order would span unequal numbers of moments along different crystallographic directions as the strength of the ordering RKKY interaction depends\cite{philip} on the interatomic separation, which is of course different in a tetragonal system along different axes. Additionally, the timescale of a neutron scattering experiment is fast enough for the neutron to see the instantaneous distribution $P(T_K)$  as if it were a static distribution. After all, neutrons are capable of measuring sound waves (phonons), and zero point motion is a superposition of phonons. Therefore, if the time-dependent distribution of Kondo temperatures in stoichiometric CeRu$_2$Si$_2$ plays a role in the low temperature response, then neutron scattering experiments should reveal short-range magnetic order whose characteristic correlation length is independent of crystallographic direction, when measured in lattice units. We show in the results section that this is indeed what has been observed.\\

Whether a time-dependent distribution of Kondo temperatures would show up in the measured specific heat and uniform susceptibility is not entirely clear, but we would expect to see it in the ac-susceptibility as we will argue in the following. We first discuss the specific heat.\\

In heavily-doped systems, when clusters separate from the main group of moments, their moments will all line up. The associated entropy shedding is reflected in the specific heat. Moreover, since the ordered environment within a cluster prevents further Kondo shielding as observed experimentally\cite{fridge}, the entropy is not recovered upon further lowering the temperature. The only degree of freedom left in the cluster, namely the overall direction of the ensuing superspin of the cluster, cannot be seen in the specific heat unless a magnetic field is applied to lift the degeneracy in this Ising system. When cluster formation follows from a time-dependent distribution of Kondo temperatures, the situation is murky. The entropy cannot be released permanently as the cluster will break up again, albeit that another fleeting cluster will appear elsewhere. Likely there will be some overall effect in the specific heat, but it should be less pronounced than for heavily-doped systems where permanent clusters appear. We would definitely expect to see a reflection of more and more moments being shielded at any given temperature, it is just unclear whether the isolated clusters would show up with an identifiable signature.\\

Similar considerations apply to the uniform susceptibility. The susceptibility reflects than an increased number of moments will be shielded, on average: this shielding shows up as an increased weakening of the average moment with decreasing temperature for all moments that are part of the lattice spanning cluster, while at the same time the overall susceptibility will display an increase upon cooling as the ratio of applied field over temperature $H/T$ increases. What is not clear is whether fleeting clusters will also show up in the susceptibility, and if so, with what signal. Most of these clusters will have a net moment (superspin) as it is unlikely that every aligned moment on the cluster will be compensated by an oppositely aligned moment. For instance, a cluster of an odd number of members will always have a net moment. However, we have to bear in mind that the cluster distribution is a dynamic one because of the underlying dynamic nature of the Kondo distribution; what becomes important is the persistance of any given cluster on the time scale of the experiment. A typical uniform susceptibility measurement involves shooting the sample through a pick-up coil on the time scale of 0.05 s. This is a long time scale compared to that of zero-point motion. We expect to see some manifestation of any clusters that form (or that materializes while the sample is making its way through the pick-up coil), but the effect will likely not be as pronounced as when permanent clusters are forced to line up with the field. As such, the saturation magnetization is expected to be lower in a stoichiometric sample than in a heavily-doped sample.\\

Perhaps the best chance of observing the signal of isolated clusters on the uniform susceptibility is by looking for a super-paramagnetic signal. Should clusters exist that happen to possess a large net moment, then the alignment of these clusters with the external field will produce a sizeable internal field as well, enhancing the susceptibility of the system. The effect should be most noticeable at the smaller fields as the alignment of a large cluster can significantly enhance the total field that other clusters experience. Thus, we would be looking for an enhancement of the susceptibility at low fields that is steeper with decreasing field than can be accounted for by a distribution of cluster sizes (and net moments).\\

While the above discussion might not necessarily sound like there is any chance of unambiguously identifying isolated clusters with a time-dependent character, ac-susceptibility measurements could reveal their presence. Suppose an isolated cluster has formed with so many dangling moments that it has a net moment of $J_{\textrm{cluster}}$. These moments can easily exceed the moment of an individual magnetic ion, as we will show in the results section. In an Ising system, this superspin has two possible orientations, and the ac-susceptibility associated with this particular cluster is given by
\begin{equation}
\chi_{ac} =\frac{(g_J J_{\textrm{cluster}}\mu_B)^2}{k_BT}\frac{1}{cosh^2(g_J J_{\textrm{cluster}}\mu_B H/k_BT)},
\label{ac}
\end{equation}
with $g_J$ the Land\'{e} g-factor which has a value of 6/7 for Ce$^{3+}$. $\chi_{ac}$ reaches a maximum value at $k_BT$ = 1.296 $g_J J_{\textrm{cluster}}\mu_B H$. Should ac-susceptibility measurements as a function of temperature and field exhibit a peak corresponding to a value of $\mu$ exceeding that of the full moment of a single magnetic ion, then this is evidence for the presence of clusters. Note that the fleeting character of such clusters does not influence the presence of the peak. If we see a peak in the susceptibility associated with a very large moment value, then this must originate from the net moment of an isolated cluster. While such large moments have not been reported, we show that low temperature ac-susceptibility measurents\cite{takahashi} on CeRu$_2$Si$_2$ do indeed reveal the presence of such large moments.\\

There is another aspect of ac-susceptibility measurements that might actually shed some light on the fleeting nature of clusters. In ac measurements, the sample is placed in a magnetic field, and a smaller, time-dependent magnetic field is superimposed on it. The induced magnetization associated with this secondary field is then recorded as a time-dependent magnetization. This time dependence is analyzed as to whether it follows the time-dependence of the secondary field (in-phase component), or whether it is out of phase with the secondary field. The in-phase component is associated with the real part of the susceptibility and the out-of-phase component with the imaginary part. For the case of fleeting clusters, we might observe an artificial out-of-phase component. When clusters spontaneously pop in and out of existence, on a timescale faster than 1/f of the ac-field, then what would be measured would simply be newly minted clusters that form and line up according to the primary field. Should this be the case, then a large imaginary component of the susceptibility would be recorded as there no longer would be full correlation between the secondary field and the (dis)appearance of the isolated clusters. In the extreme case, the experiments would record that the real and imaginary parts would be equal, indicating that the clusters do not exhibit any persistance on the timescale of the experiment. We will show that ac-susceptibility experiments on CeRu$_2$Si$_2$ do indeed show that this could well be the case.\\

\section{results}
Heavily doped Ce(Ru$_{0.755}$Fe$_{0.245}$)$_2$Ge$_2$ and stoichiometric CeRu$_2$Si$_2$ have much in common. Both systems display a strong Ising character (the tetragonal c-axis being the easy axis for the moments of the Ce-ions) because of the crystal electric fields splitting the $J_z$-energy levels into three doublets.\cite{isingnature} The first excited state in CeRu$_2$Si$_2$ is well separated\cite{isingnature} from the lowest level ($\Delta$ $>$ 30 meV), and neutron scattering experiments\cite{flouquet,kadowaki} demonstrated that the ordered moments point along the c-axis. The ground state mostly has a $J_z$ = 5/2 character.\cite{isingnature} Both systems display short-range magnetic correlations associated with incipient spin density wave (SDW) order.\cite{flouquet,kadowaki} Substitution of about 3.5 \% rhodium on the Ru-sites drives CeRu$_2$Si$_2$ into a long-range ordered SDW ground state.\cite{kadowaki2,lro} Thus, quantum critical disordered Ce(Ru$_{0.755}$Fe$_{0.245}$)$_2$Ge$_2$ can be directly compared to stoichiometric CeRu$_2$Si$_2$ and nearly stoichiometric Ce(Ru$_{0.97}$Rh$_{0.03}$)$_2$Si$_2$ in order to test whether spontaneous magnetic clusters can form in stoichiometric systems.\\

\subsection{Neutron scattering}
In this section, we restrict our comparison to CeRu$_2$Si$_2$ as this was the only system we found in the literature where the magnetic correlation lengths had been measured along independent crystallographic directions. Neutron scattering experiments in which the magnetic correlation lengths associated with short-range order are measured along independent crystallographic directions provide direct insight into the ordering mechanism. The work of Flouquet {\it et al.}\cite{flouquet} showed that CeRu$_2$Si$_2$ is on the cusp of a spin density wave transition. This was confirmed in inelastic neutron scattering experiments on large single crystals by Kadowaki {\it et al.}\cite{kadowaki}  and in 2006\cite{kadowaki2} on the lightly-doped Ce(Ru$_{0.97}$Rh$_{0.03}$)$_2$Si$_2$ compound that was on the verge of long-range order. Similarly, the tell-tale SDW incommensurate ordering wavevectors were observed\cite{wouterprb} in heavily-doped Ce(Ru$_{0.755}$Fe$_{0.245}$)$_2$Ge$_2$. Both the heavily-doped and the lightly-doped compounds appear to be at comparable points in their phase diagrams: the paramagnetic phase on the verge of long-range ordering.\\

\begin{figure}[b]
    \begin{center}
    \includegraphics*[viewport =75 55 580 550,width =85mm,clip]{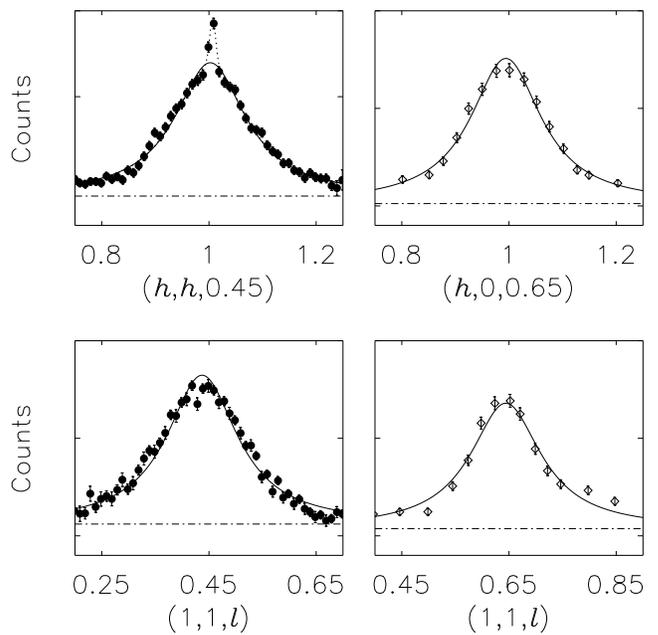}
    \end{center}
    \caption{Elastic neutron scattering data for Ce(Ru$_{0.755}$Fe$_{0.245}$)$_2$Ge$_2$\cite{wouterprb} at T  = 2K (left two panels) and for E = 0.2 meV for Ce(Ru$_{0.97}$Rh$_{0.03}$)$_2$Si$_2$\cite{kadowaki2} at T = 1.6 K as a function of momentum transfer along the directions specified on the horizontal axes. The solid line through all data is a Lorentzian fit with width of 0.166/2 rlu for all four panels. } 
    \label{direct}
\end{figure}

Long range order has been observed\cite{wouterprb,lro} in both systems when chemically doped a fraction more into the SDW-range.  In Fig. \ref{direct}, we show elastic neutron scattering data on single crystal Ce(Ru$_{0.755}$Fe$_{0.245}$)$_2$Ge$_2$ where one part of the crystal ended up being slightly overdoped,\cite{wouterprb,fridge} resulting in a spectrometer resolution limited Bragg peak. Masking this part of the crystal removed the sharp peak, indicative of long-range order, leaving the scattering associated with short-range order (see left panels in Fig. \ref{direct}). Experiments on Ce(Ru$_{1-x}$Rh$_{x}$)$_2$Si$_2$ with slighly higher rhodium concentrations also revealed the existence\cite{lro} of a fully developed SDW ground state. Thus, Ce(Ru$_{0.755}$Fe$_{0.245}$)$_2$Ge$_2$ and Ce(Ru$_{0.97}$Rh$_{0.03}$)$_2$Si$_2$ appear to be at identical points in their Doniach\cite{doniach} phase diagram.\\

Given that the ordered phase in both compounds has been identified unambiguously\cite{kadowaki2,fridge} as a SDW, we know that the ordering interaction must be the RKKY interaction. This interaction depends on the separation between magnetic ions in an oscillatory manner, with the period of oscillation determined by the Fermi wave vector and the strength varying as $\sim 1/r^4$. Thus, the strength of interaction not only differs in plane and along the body diagonal in these tetragonal compounds, but it also differs between the two compounds: $d_{Ce-Ce}$ along a-axis is 4.08 {\AA} in Ce(Ru$_{0.755}$Fe$_{0.245}$)$_2$Ge$_2$ and 4.20 {\AA} in CeRu$_2$Si$_2$, and  along the body diagonal we have $d_{Ce-Ce}$ = 5.97 {\AA} in Ce(Ru$_{0.755}$Fe$_{0.245}$)$_2$Ge$_2$ and 5.73 {\AA} in CeRu$_2$Si$_2$. Therefore, unless clusters materialize as we explain below, short-range order should span different numbers of moments along different crystallographic directions, reflecting the strength of the ordering interaction along those directions.\\

When clusters appear, because the distribution of Kondo temperatures has led to groups of moments being isolated from the rest of the lattice when surrounded by shielded moments\cite{heitmannconf,heitmannonline,fridge}, then the moments on the clusters all line up with their neighbors because of quantum mechanical finite-size effects\cite{fridge}. In this case, the strength of the ordering interaction no longer plays a role. Instead, we observe that all moments on a cluster are aligned, and the measured correlation lengths reflect the cluster sizes. When clusters form because of random processes, such as doping or zero-point motion, then all correlation lengths will span equal numbers of moments along any crystallographic direction. This was indeed what was observed in heavily-doped Ce(Ru$_{0.755}$Fe$_{0.245}$)$_2$Ge$_2$\cite{wouterprb,fridge} and is shown in the left panels of Fig. \ref{direct} where the observed short-range scattering along the c-direction and along the (110)-direction can be seen to span identical numbers of moments.\\

Kadowaki {\it et al.} have performed similar experiments\cite{kadowaki,kadowaki2} on CeRu$_2$Si$_2$ and on Ce(Ru$_{0.97}$Rh$_{0.03}$)$_2$Si$_2$. We reproduce their results on quantum critical Ce(Ru$_{0.97}$Rh$_{0.03}$)$_2$Si$_2$ in Fig. \ref{direct} for an energy transfer of 0.2 meV. Thus, the two compounds might not be entirely directly comparable as the data on Ce(Ru$_{0.755}$Fe$_{0.245}$)$_2$Ge$_2$ was collected at $E$ = 0 meV with a looser energy collimation than the one employed by Kadowaki. However, both data sets are free from non-magnetic scattering and were measured at comparable temperatures. We observe in the right hand panels that the data on Ce(Ru$_{0.97}$Rh$_{0.03}$)$_2$Si$_2$ also demonstrate that the correlation lengths of the incipient order span equal number of moments along the a- and c-direction. Moreover, the solid curve through all data sets for both compounds is a Lorentzian curve centered at the ordering wave vector with full width of 0.166 rlu (reciprocal lattice units). As can be seen, this curve provides a good description of all four data sets.\\

 We are not aware of any explanation other than ordered clusters that could explain short-range order that, on the one hand, is associated with an interaction whose strength depends on separation between the magnetic moments, but, on the other hand,  whose spatial extent does not depend on the intermoment separations. In addition, when we include the different ratios $d_{Ce-Ce}^{diagonal}$/$d_{Ce-Ce}^{a}$ for the two compounds (1.463 and 1.365 for Ce(Ru$_{0.755}$Fe$_{0.245}$)$_2$Ge$_2$ and Ce(Ru$_{0.97}$Rh$_{0.03}$)$_2$Si$_2$, respectively), then we can also rule out an accidental effect intrinsic to the RKKY interaction. Lastly, when we compare the data taken by Kadowaki at higher temperatures (Fig. 1 in reference [\onlinecite{kadowaki2}]), it appears that within the accuracy of the experiments the equivalence of the correlation lengths between the a and c-directions continues to hold at all temperatures. This was also observed in Ce(Ru$_{0.755}$Fe$_{0.245}$)$_2$Ge$_2$\cite{fridge} and it shows that all scattering at the incipient ordering vector is associated with clusters.\\ 

In conclusion, neutron scattering experiments offer strong evidence for the presence of magnetic clusters in (near) stoichiometric quantum critical systems. As discussed in the preceding section, neutron scattering experiments cannot reveal whether these clusters are fleeting or permanent as the speed of a neutron is comparable to the speed of ionic motion.\\
  
\subsection{Specific heat and uniform (dc) susceptibility}
When we have a static distribution of Kondo shielding temperatures, percolation theory\cite{stauffer} tells us what to expect for the occupation dependence of the various quantities. When we have a dynamic distribution, we are in unchartered territory. Our approach in this section is to review the results for a static Kondo distribution pertinent to quantum critical Ce(Ru$_{0.755}$Fe$_{0.245}$)$_2$Ge$_2$. Then we will compare the results for Ce(Ru$_{0.755}$Fe$_{0.245}$)$_2$Ge$_2$ with those of CeRu$_2$Si$_2$ and Ce(Ru$_{0.97}$Rh$_{0.03}$)$_2$Si$_2$ with the aim of identifying the part of the response caused by fleeting clusters.\\

When we have a static distribution of Kondo shielding temperatures then moments are only allowed to be removed from the lattice spanning collection of moments as the moments of the isolated clusters are protected from Kondo shielding because of their ordered environment which severely impeded the Kondo shielding process\cite{philip,fridge}. Under these conditions there exists a direct connection between the strength of the lattice spanning cluster (how many moments it consists of) and the specific heat of the system. Whenever a moment is removed (shielded) in an Ising systsem, then $k_B$ln2 in entropy is removed. When a moment is removed that results in a cluster of $s$ sites separating from the lattice spanning cluster, then $sk_b$ln2 in entropy is removed. This consists of the one moment that was shielded, and the loss in entropy of $(s-1)k_B$ln2 for the cluster as all the moments in this cluster will line up, leaving only one degree of freedom for the superspin of the cluster. But note that this superspin degree of freedom will not be released, unless the system is placed in a magnetic field. Thus, when $s$ sites peel off, the system effectively loses all entropy associated wth $s+1$ sites. This is the same statement as saying that all the available entropy of the system is locked up in the lattice spanning cluster. The demise of the entropy is in lockstep with the demise of the infinite cluster.\\

The correspondence between the entropy and the infinite cluster allows one to perform (site) percolation computer simulations as a function of occupancy, and then use this correspondence to convert occupancy to occupancy as a function of temperature. This in turn then allows other simulated properties, such as correlation length, to be compared between computer simulations and experiment. This was first tried in reference [\onlinecite{gaddy}]. However, there is a more direct way to relate the measured specific heat to the observed uniform susceptibility using this correspondence, bypassing the in-between computer simulations.\\

When the conduction electrons Kondo shield the local moments, it is not a process that takes place abruptly, rather it is spread out over 4 decades in $T/T_K$ with most of the shielding taking place around $T\sim T_K$. The same applies to the susceptibility associated with the local Ce-moment when it transitions from a free moment to Kondo singlet. Bearing in mind the Ising nature of the Ce-moment, we can make the following numerical association to arrive at the susceptibility of the infinite cluster $\chi_{\infty}$ while its moments are undergoing shielding:
\begin{equation}
\chi_{\infty}(T)  = \frac{S(T)}{\textrm{R ln2}}\chi_{\textrm{free}}(T).
\label{inf}
\end{equation}
In here, $S(T)$ is the experimentally determined entropy of the system, inferred by numerically integrating the specific heat $c(T)/T$. R is the gas constant and $\chi_{\textrm{free}}$ is the known susceptibility of an individual Ce-ion placed in the local tetragonal crystal electric field (CEF). From experiments on CeRu$_2$Si$_2$, we know\cite{isingnature} that the field levels are 33 and 55 meV and therefore, barely influence the specific heat below 100 K. We do not have values for the CEF levels in Ce(Ru$_{0.755}$Fe$_{0.245}$)$_2$Ge$_2$, but based on the Ising nature of the moments observed in neutron scattering experiments,\cite{wouterprb} we assume that they are similarly valued. Therefore, we chose the factor Rln2 to normalize the measured entropy to the fraction of surviving moments as this is the total amount of entropy that will be shed when a mol of Ising spins loses their freedom to point up or down.\\

\begin{figure}[tb]
    \begin{center}
    \includegraphics*[viewport =120 100 580 450,width =85mm,clip]{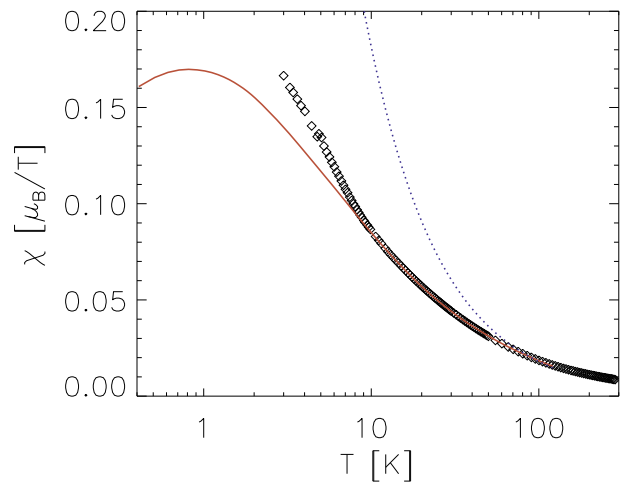}
    \end{center}
    \caption{The uniform susceptibility for Ce(Ru$_{0.755}$Fe$_{0.245}$)$_2$Ge$_2$ (diamonds) measured\cite{fridge} in a field of 0.2 T parallel to the crystallographic c-axis in this (almost) Ising system. The dotted curve follows the Curie Weiss law at high temperatures which has only a very limited range of validity. The solid curve (see Eq. \ref{inf}) is the measured entropy of the sample, divided by Rln2 and multiplied by the temperature dependence of the magnetic response of a three level doublet system with energy gaps of 33 and 55 meV,\cite{isingnature} with the ground state magnetic moment taken to be 1.43 $\mu_B$ (and the two excited state moments reduced by a similar factor, although the excited states do not play a factor in the visual comaprison).  The measured susceptibility and the specific-heat based prediction start to deviate around T $<$ 10 K where clusters first appear in the neutron scattering spectra.} 
    \label{uniform}
\end{figure}

There is still one adjustable parameter left in Eq. \ref{inf}, namely the size of the moment for the free Ce-ion. Based on the admixture of the $|J_z =5/2\rangle$ and  $|J_z =3/2\rangle$ states, we can expect a free moment of around 1.85 $\mu_B$ (using $g_J$ =6/7). We find that this appears to be an overestimation, and instead, we used a value of 1.43$\mu_B$ in our Eq. \ref{inf} based on the comparison shown in Fig. \ref{uniform}, a value picked to obtain the best visual agreement. With this reduction of the free moment (in the process of being Kondo shielded), the agreement predicted by Eq. \ref{inf} is perfect between 10 and 120 K (with 120 K the upper bound of our measured specific heat data. Encouragingly, Eq. \ref{inf} starts to fail below 10 K, which is exactly where we would expect it to fail. Below 10 K, neutron scattering experiments showed\cite{wouterprb} the first appearance of isolated clusters, with more and larger clusters appearing with decreasing temperature. When those ordered clusters have a net moment, then they contribute to the susceptibility since their superspin can align with the external magnetic field. In conclusion, it appears that Eq. \ref{inf} correctly describes the susceptibility of the disordered moments and that it is a useful tool when looking for the appearance of magnetic clusters with a static distribution.\\ 

The situation in CeRu$_2$Si$_2$ and Ce(Ru$_{0.97}$Rh$_{0.03}$)$_2$Si$_2$ is not as clear-cut: Eq. \ref{inf} works very well at the lowest temperature where the magnetic specific heat curve $C(T)/T$ as well as the uniform susceptibility were found to be T-independent in moderate fields ($H$  = 0.1 T) and low temperatures (1.5 $<$ $T$ $<$ 5 K). However, Eq. \ref{inf}  does not give a good description for the susceptibility in the intermediate temperature range, even when we allow for the vastly different values reported for $c/T$. We discuss this in the following.\\ 

\begin{figure}[tb]
    \begin{center}
    \includegraphics*[viewport =180 160 590 620,width =85mm,clip]{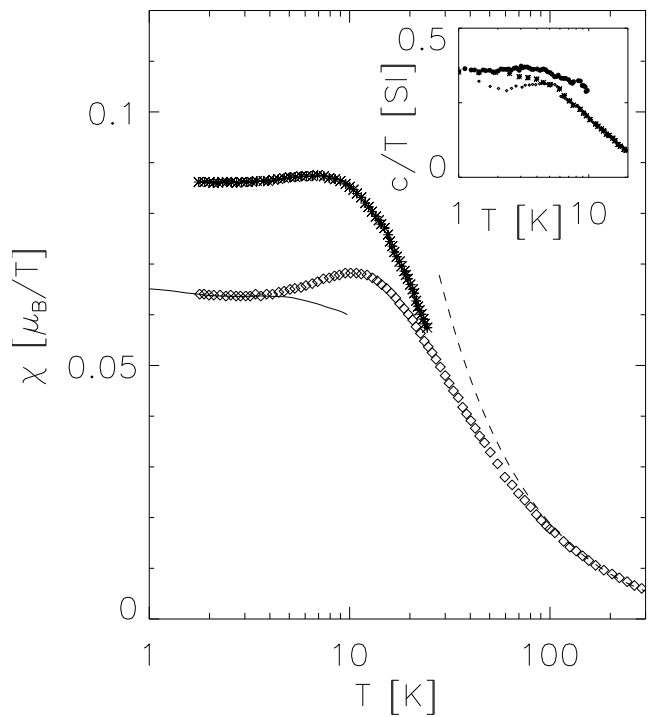}
    \end{center}
    \caption{The uniform susceptibility for  CeRu$_2$Si$_2$ (diamonds) Ce(Ru$_{0.97}$Rh$_{0.03}$)$_2$Si$_2$ (stars) measured for a field of 0.1 T parallel to the c-axis\cite{tabata}. All the data in the figure are scanned in by hand and may not exactly coincide with the published data points. The solid curve (see Eq. \ref{inf}) is the measured entropy of the sample using the data by Tabata\cite{tabata} for $c/T$, divided by Rln2 and multiplied by the temperature dependence of the magnetic response of a three level doublet system with energy gaps of 33 and 55 meV,\cite{isingnature} with the ground state magnetic moment taken to be 1.21 $\mu_B$. The measured susceptibility and the specific-heat based prediction start to deviate around $T$ = 5 K, with the prediction underestimating the measured susceptibility. The dashed curved is the high temperature susceptibility using the same energy gaps, but now with a ground state moment of 1.68 $\mu_B$. The inset shows the $c/T$ data for  CeRu$_2$Si$_2$ measured by three groups: Laquerda {\it et al.}\cite{visser} (diamonds), Besnus {\it et al.}\cite{besnus} (stars), and Tabata\cite{tabata} (circles).} 
    \label{susstoichio}
\end{figure}

The specific heat curve below $T$ = 5 K for CeRu$_2$Si$_2$ is virtually $T$-independent \cite{tabata,besnus,visser} with $\gamma  = c/T$  = 0.38 J/mol/K$^2$. The value for Ce(Ru$_{0.97}$Rh$_{0.03}$)$_2$Si$_2$ is slightly higher with a weak temperature dependence\cite{tabata} ($\gamma$ = 0.5 J/mol/K$^2$ at $T$ = 0.2 K, and $\gamma$ = 0.4 J/mol/K$^2$ at $ T$ = 5 K). A problem with the reported values for the specific heat is that the three sources start to differ by as much as 50\% at $T$ = 10 K. We show this discrepancy in the inset of Fig. \ref{susstoichio}. In here, we use the data reported by Tabata\cite{tabata} since the measurements for the susceptibility on the same samples were reported in the same publication. We note that our observation, that Eq. \ref{inf} gives a good description at the lowest temperatures but {\it not} at intermediate temperatures is not altered when we take into account the discrepancies in reported specific heat curves.\\

When $c/T$ equals a constant at low temperatures, then the entropy of the system depends on temperature as $S =\gamma T$. When we are in the region where $H/T\ll 1$ then we can approximate the non-interacting susceptibility as $\chi_{free}$ = $(g_J J_{\textrm{eff}} \mu_B)^2/k_B T $ and Eq. \ref{inf} reduces to
\begin{equation}\label{lowt}
    \begin{split}
        \chi_{\infty}(T)  &= \frac{\gamma}{\textrm{R ln2}}\frac{(g_J J_{\textrm{eff}} \mu_B)^2}{k_B} \\
 &= 0.1166 \gamma (g_J J_{\textrm{eff}} )^2 \frac{ \mu_B}{\textrm{T.Ce-ion}} 
    \end{split}
\end{equation}
Thus, Eq. \ref{inf} predicts a $T$-independent susceptibility when $c/T$ is temperature independent. Applying Eq. \ref{lowt} to the measured values, we deduce a low temperature moment for the Ce-ions that form the lattice spanning cluster of 1.21 $\mu_B$/Ce for both CeRu$_2$Si$_2$ and Ce(Ru$_{0.97}$Rh$_{0.03}$)$_2$Si$_2$. This value appears entirely reasonable when compared to the high-temperature susceptibility (Fig. \ref{susstoichio}) of 1.65 $\mu_B$/Ce-ion when fitting to a system that is a single doublet (and  2.33 $\mu_B$/Ce-ion when fitting to a pair of doublets). Moreover, even though the values for $\chi$ differ by 30\% between the two compositions, the low temperature unshielded moment value remains identical, as should be expected upon minimal doping.\\

Applying Eq. \ref{inf} to the intermediate temperature range (5 $<$ $T$ $<$ 30 K) reveals clear deviations, even when taking into account the variation in specific heat curves. The broad bump in the susceptibility near $T \sim$ 10 K is not reproduced by applying Eq. \ref{inf}. We saw a similar discrepancy in  Ce(Ru$_{0.755}$Fe$_{0.245}$)$_2$Ge$_2$ when isolated clusters formed with a superspin. Assuming that we are seeing similar behavior in CeRu$_2$Si$_2$, we would have to conclude that in this system with an average Kondo temperature\cite{tabata} of 24 K that small clusters start to form in the intermediate temperature region; however, when the temperature is lowered further and the Kondo interaction becomes stronger\cite{wilson}, then apparently these smaller clusters can still be shielded. This in contrast to the case for  Ce(Ru$_{0.755}$Fe$_{0.245}$)$_2$Ge$_2$ (with a static Kondo distribution and an average Kondo temperature of $\sim$ 15 K) where we have not seen evidence of smaller clusters being shielded upon lowering the temperature. We cannot ascertain at this point whether the above is a reasonable interpretation for the discrepancy, or not, although it is known from inelastic neutron scattering experiments (Fig. 2 in reference [\onlinecite{kadowaki}]) on CeRu$_2$Si$_2$ that increased scattering intensity appears at the SDW ordering wave-vector in the intermediate temperature range starting at $T$ $\sim$ 20 - 30 K, indicating that clusters start to form around $T_K$.\\

The uniform susceptibility provides another strong hint for the presence of ordered, magnetic clusters. It has been observed\cite{paramagnetic} in CeRu$_2$Si$_2$ that the susceptibility at very low fields is not linear in magnetic fields but increases rapidly with decreasing magnetic field. This is what is expected when ferromagnetic impurities are present in a sample as the molecular fields created by these impurities amplify the total field experienced by other moments when they line up with the external fields. This results in a measured response that is no longer linear in the applied field for small fields. The level of impurities required to explain the super-paramagnetic behavior in CeRu$_2$Si$_2$ was calculated\cite{paramagnetic} to be of the order of 10$^{-3}$ $\mu_B$ per Ce-ion. However, this level of impurities appears to be rather high considering that starting materials are of a 5N purity. We argue that this non-linear response is instead due to the presence of clusters.\\

\begin{figure}[tb]
    \begin{center}
    \includegraphics*[viewport =120 155 520 570,width =85mm,clip]{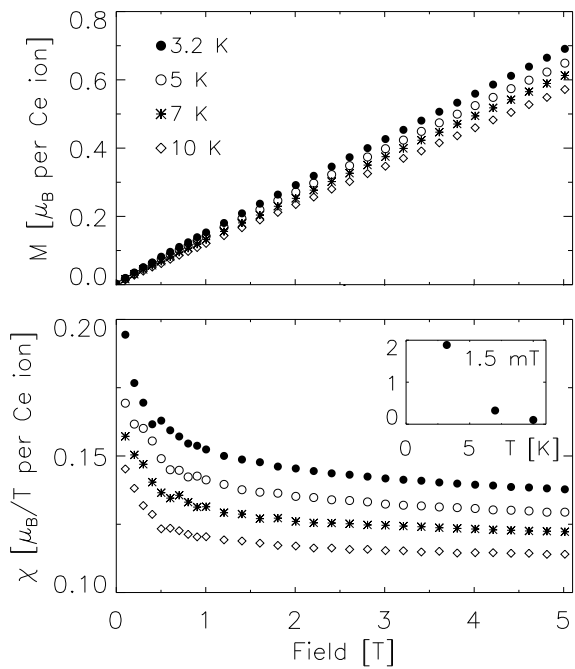}
    \end{center}
    \caption{(top panel) Magnetization data\cite{fridge} for Ce(Ru$_{0.755}$Fe$_{0.245}$)$_2$Ge$_2$ as a function of applied field along the c-axis for the temperatures indicated in the figure. The bottom panel displays the accompanying susceptibility data. Note the upturn at fields below H  = 1 T. The inset shows the temperature dependence of the susceptibility, on the same scale, in a very small applied field of 1.5 mT. } 
    \label{fig3}
\end{figure}

\begin{figure}[tb]
    \begin{center}
    \includegraphics*[viewport =170 140 580 420,width =85mm,clip]{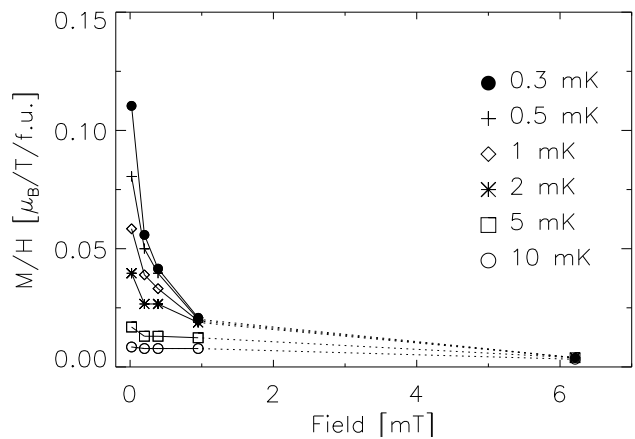}
    \end{center}
    \caption{The susceptibility as a function of applied field a few tens of degrees away from the c-direction\cite{takahashi} for CeRu$_2$Si$_2$ at very low temperatures. The data have been normalized to the observed paramagnetic, temperature independent susceptibility measured\cite{takahashi} for $T$ $>$ 50 mK, that is, the zero on the vertical axis corresponds to this paramagnetic level. The sharp upturn in the susceptibility below $H$  = 1 mT is quite pronounced, especially as the temperature is lowered below $T$ = 3 mK.} 
    \label{fig4}
\end{figure}

When clusters are present at low temperatures, some of these clusters will have a large net moment. In fact, this net moment is on average of the order of 2 - 3 $\mu_B$ as we show in the next section, with some rare clusters have a very large moment. Such clusters take on the role of ferromagnetic entities, capable of locally enhancing the applied external magnetic field. We show in Figs. \ref{fig3} and \ref{fig4} that this super-paramagnetic response is also present in  Ce(Ru$_{0.755}$Fe$_{0.245}$)$_2$Ge$_2$\cite{fridge} measured at low temperatures and in high-purity CeRu$_2$Si$_2$ measured\cite{takahashi} at very low temperatures and very low fields. We also mention that Tabata\cite{tabata} observed this behavior in CeRuRhSi$_2$ and observed that it could not be satisfactorily explained by modeling it with the Kondo disorder model. Note, the Kondo disorder model does not include the formation of clusters.\\

Based on the prevalence of the observation of this super-paramagnetic effect in four different samples, we believe it is much more likely that we are actually seeing a feature of the system's response rather than a manifestation of unintended contamination at a level exceeding the purity of the starting ingredients. We can also turn the reasoning around: should clusters be present, then percolation theory\cite{stauffer} tells us that there must be a large number of clusters with a net moment, and a small number of clusters with a very large net moment. Therefore, for the cluster model to be valid, we must observe a super-paramagnetic response.  While the presence of this effect is not necessarily proof of the presence of clusters, an absence of this effect would have implied an absence of (large) ordered clusters.\\

In summary, the uniform susceptibility data do not yield direct evidence for our cluster interpretation. The link between a temperature independent $c/T$ curve and a temperature independent susceptibility is strong, and the ensuing free moment values are in line with expectations and agree between the stoichiometric compound and the lightly doped compound, even when the underlying curves differ by 30\%. We would expect to see isolated clusters with superspins appear around the average Kondo temperature, but unlike the case for heavily-doped  Ce(Ru$_{0.755}$Fe$_{0.245}$)$_2$Ge$_2$, it appears that the smaller clusters can be reabsorbed, presumably when the dynamic Kondo distribution creates a new set of clusters, opening moments up to shielding again. The cluster scenario does offer a very natural explanation for the super-paramagnetic behavior observed in the 122-systems, even in compounds that have been prepared using a purity of starting materials higher than the observed ferromagnetic effects should it be attributable to impurities. In the next subsection we show that ac-susceptibility measurements at very low temperatures do produce firm evidence for the existence of clusters in stoichiometric CeRu$_2$Si$_2$.\\

\subsection{Ac susceptibility}
Takahashi {\it et al.}\cite{takahashi} performed a series of magnetization and ac-susceptibility measurements on CeRu$_2$Si$_2$ at very low temperatures (milliKelvin) and fields (milliTesla). They found considerable temperature and field dependence of the susceptibility in this range where earlier experiments in larger fields (0.1 - 0.2 T) in a higher temperature range ($T$ $>$ 2 K) had found\cite{flouquet} that the uniform susceptbility would reach a constant value for $T$ $<$ 10 K (see Fig. \ref{fig3}), although this value was strongly dependent of the rhodium concentration.\cite{tabata} The authors determined two vastly differing Ce-moment values from their data: based on the region of low $\mu_BH/k_BT$-values, they deduced a moment value of 0.01 $\mu_B$/Ce-ion from a Curie-behavior fit to the susceptibility; they found a much lower moment (by a factor of about 1,000) based on the magnetization at high $\mu_BH/k_BT$-values where the magnetization became independent of $H/T$.\\

In fact, there is a third moment scale present in the data\cite{takahashi} by Takahashi {\it et al.}, indicative of an unusually large moment. The authors noticed that the ac-susceptibility showed a distinct maximum in fields of 0.2, 0.39, and 0.94 mT. The peak position occurred at $T$ = 0.5, 0.9, and 3 mK, respectively.  Since CeRu$_2$Si$_2$ is an Ising system at low temperature, we can use Eq. \ref{ac} and its predicted peak position at  $k_BT$ = 1.296 $g_J\mu H$ to associate this peak with a moment value of 3 $\pm$ 0.5 $\mu_B$. This value exceeds the moment value for an unshielded Ce-ion. While the authors did not discuss the moment value associated with the peak in the ac-susceptibility, they did rule out\cite{takahashi} small amounts of disorder and a putative spin glass phase as the cause of the factor 1,000 discrepancy between susceptibility and magnetization inferred Ce-moment values.\\

We argue that these vastly differing moment values can be qualitatively understood assuming that clusters do indeed materialize in CeRu$_2$Si$_2$ at low temperatures, and that the moments on clusters line up with their neighbors because of finite-size effects. For the sake of simplicity in making our arguments, we assume that the moments on isolated clusters order anti-ferromagnetically such that each unit cell of CeRu$_2$Si$_2$ containing two formula units has a net moment of zero when it houses two surviving (unshielded) anti-ferromagnetically ordered Ce-moments. This simplification makes it easier to count as to whether a cluster ends up with a net moment or not, and how many uncompensated moments persist. It is the sum of these uncompensated moments that give the cluster a net moment, or superspin. In an Ising system the up/down orientation of this superspin can still be influenced by an external magnetic field. In Ce(Ru$_{0.755}$Fe$_{0.245}$)$_2$Ge$_2$, this reorientation degree of freedom was identified\cite{jap} as the low-energy excitations responsible for the observed $E/T$-scaling.\\ 

\begin{figure}[tb]
    \begin{center}
    \includegraphics*[viewport =180 140 580 420,width =85mm,clip]{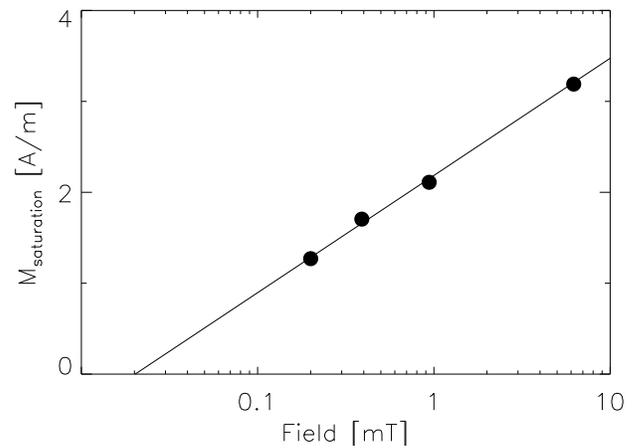}
    \end{center}
    \caption{The saturation magnetization for CeRu$_2$Si$_2$ measured\cite{takahashi} in low fields and plotted as a function of the logarithm of applied field.  Fitting the data to a log-dependence, we find that the saturation magnetization reaches zero at $H$ = 0.02 mT.} 
    \label{saturation}
    \end{figure}

Before we discuss our cluster interpretation, we note that in their paper Takahashi {\it et al.} referred\cite{takahashi} to the magnetization level where the value became independent of $H/T$ as saturation magnetization and used it to determine a saturation moment of $\sim$ 10$^{-5}$ $\mu_B$/Ce-ion. While the magnetization level clearly became independent of $H/T$ for values above $\sim$ 1 T/K, the actual level was still seen to depend of the applied magnetic field (see Fig. \ref{saturation}). As such, it is not a saturation magnetization as is normally understood when we discuss curves such as the Brillouin function. Moreover, the interpretation of a moment of 10$^{-5}$ $\mu_B$/Ce-ion is not consistent with the interpretation of this being a saturation level. For moments that small, saturation should only be reached for $H/T$ values of around 10$^5$ T/K, not 1 T/K. One final note is that the data were presented with reference to the paramagnetic level observed in the susceptibility for $T$ $>$ 50 mK. It was unclear whether the authors referred to the level of 0.03 emu/mol as measured in much higher fields for $T$ $>$ 1 K\cite{flouquet,tabata} (see Fig. \ref{fig3}), or whether their paramagnetic level used as an offset was different.\\

Based on our understanding of cluster formation caused by a permament distribution of Kondo temperatures, we propose the following for stoichiometric CeRu$_2$Si$_2$ based on a dynamic distribution of Kondo temperatures. At any moment in time, CeRu$_2$Si$_2$ is subject to a distribution of Kondo temperatures. The Kondo temperature at any given Ce-site dictates whether that Ce-moment will be (mostly) shielded, or persist as an unshielded moment. Therefore, at any temperature we can expect manifestations of percolation physics, and at low temperaure we can expect the appearance of isolated clusters. As in Ce(Ru$_{0.755}$Fe$_{0.245}$)$_2$Ge$_2$, the moments on isolated clusters have to line up with their neighbors because of finite-size effects, provided that the temperature is low compared to the allowed quantized energies of disordering spin fluctuations on the cluster.\\

An isolated cluster can acquire a net overall moment. When the moments line up, all pairs of neighboring moments will cancel each other out, but there will be dangling moments on the surface of the cluster. Of course, surface is a term we use loosely here as the topology of a cluster that formed by random removal of moments is closer to that of a fractal than that of a solid object.\cite{stauffer} When we take susceptibility measurements, we only see the net moments of the clusters. The net moment only represents a small fraction of all the moments on the cluster given the anti-ferromagnetic ordering. Also, only a small fraction of the moments will end up in clusters. For reference, for quantum critical Ce(Ru$_{0.755}$Fe$_{0.245}$)$_2$Ge$_2$, 27\% of the moments are believed to end up in clusters at the QCP. This number corresponds to the percolation threshold for protected percolation.\cite{fayfar} For large clusters, somewhere in the neighborhood of 1\% of the moments may end up not being compensated, as we detail below. The main consequence of cluster formation is that the entities that give rise to the susceptibility signal are far less numerous than the number of Ce-ions in our sample: there are far fewer clusters than there are unit cells, and there are far fewer uncompensated moments on a cluster than there are compensated moments on a cluster.\\

We illustrate how these considerations lead to an artificially large ratio between the moment determined from the high temperature susceptibility compared to that determined from the saturation magnetization. When we determine the moment per Ce-ion from the high temperature susceptibility, we compare our measured signal to $N\mu^2$, with N the known number of Ce-ions in our sample. Similarly, for determining the moment based on the saturation magnetization $\mu_s$, we compare our signal to $N\mu_s$. We then simply divide both terms by $N$ so that we can compare the two moments; provided we have reached saturation, the two moment values should be identical. However, when only a fraction $f$ of all the moments end up being the uncompensated cluster moments that produce the measured susceptibility and magnetization, then our normalization is wrong by a factor of $f$, and after taking the square root of $N\mu^2$, the factor $f$ does not cancel, but we end up with an overestimation of $\mu$ based on the high temperature susceptibility by a factor of $1/\sqrt{f}$. From computer simulations we estimate $f$ to be in the range of 0.1 - 2\%, accounting for a factor of 5-30 in the observed\cite{takahashi} discrepancy of $\sim$ 1,000 discussed above.\\

There are three more factors that reduce the observed discrepancy. First, the authors\cite{takahashi} assumed that the high temperature term in the susceptibility would be proportional to $\mu^2/3k_B$, but for an Ising system, the proportionality is given by $\mu^2/k_B$, removing a factor of $\sqrt{3}$. Second, the level of discrepancy is dependent on the applied field (see Fig. \ref{saturation}), with the highest field in the study ($H$ = 6.2 mT) yielding the smallest discrepancy factor of 720. Using the data in Fig. \ref{saturation}, the discrepancy would be reduced by another factor of 2 by the time the applied field reaches 0.1 T. Third, when there is a distribution of net moments, then the moment derived from the high temperature susceptibility will exceed that of the saturation magnetization. This is a result of averaging over $\mu^2_{\textrm{cluster}}$ verus averaging over $\mu_{\textrm{cluster}}$. Large net moments will skew the $\mu^2_{\textrm{cluster}}$ averaging towards higher values. For example, imagine applying the averaging procedure to a system of 101 clusters, one with a net moment of 100, and 100 clusters with a net moment of 1. With these numbers the moment determined from the saturation magnetization would be 2, but the one determined from the susceptibility would be 10. This illustrates how large clusters can skew what is measured in the high temperature susceptibility. When we combine all these factors, we see that the discrepancy is greatly reduced. Given the uncertainties, we cannot tell whether it will disappear completely, but from computer simulations to be discussed below, we estimate that the discrepancy is reduced to within a factor of 5-10.\\

The appeareance of clusters with a superspin leads to a very natural explanation for the high-moment peak in the ac-susceptibility, as well as the overall small level for the average moment. In order to clarify this statement, we have performed a site percolation computer simulation on a lattice of 400 x 400 x 400 magnetic sites using a body-centered nearest-neighbor topology. In the simulations, sites were removed at random until the percolation threshold was reached. We added the restriction that sites could not be removed from isolated clusters. This protected type of percolation\cite{fayfar} is the one that describes the clusters that form\cite{fridge} in Ce(Ru$_{0.755}$Fe$_{0.245}$)$_2$Ge$_2$, which is why we will also use this method for comparison to body centered CeRu$_2$Si$_2$. Of course, a fleeting distribution of Kondo temperatures is not the same as a permanent distribution, nor is stoichiometric CeRu$_2$Si$_2$ truly quantum critical; for that we need\cite{tabata} 3.5\% Rh-doping. Notwithstanding, using the cluster distribution at the percolation threshold generated through random moment removal while leaving isolated clusters intact is expected to shed light on how many moments become dangling moments, how many clusters we can expect to appear that have a superspin, and what the average net moment is associated with such superspins.\\

\begin{figure}[tb]
    \begin{center}
    \includegraphics*[viewport =120 100 580 560,width =85mm,clip]{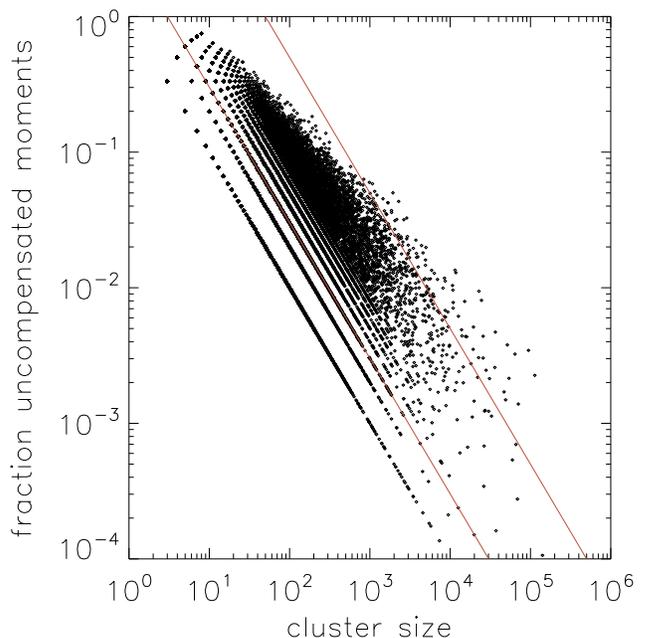}
    \end{center}
    \caption{The distribution of uncompensated moments as a function of cluster size as determined from a computer simulation at the percolation threshold. Clusters of size one and clusters with a net moment of zero have been omitted. The two solid lines correspond to an average number of uncompensated moments per cluster of 3 (lower line), and to a high superspin of 50 uncompensated moments (upper curve).} 
    \label{clusdist}
\end{figure}

400$^3$ sites were simulated whose positions were characterized by 3 integer coordinates. The individual moment direction (up/down) was based on whether the z-direction integer was odd or even, representing an Ising system with anti-ferromagnetic interactions. Moments were removed at random from the collection of moments that span the lattice (the so-called lattice spanning, percolating, or infinite cluster) until the last connection was severed so that the cluster would no longer extend from one side of the lattice to the other. Also, we used periodic boundary conditions to ensure we would not suffer from edge effects such as large artificial net moments for clusters terminating on an edge. The step before the threshold was reached, we took inventory of all remaining clusters and tabulated their size (how many moments there were per cluster) and their net moment. This is shown in Fig. \ref{clusdist}. We will use this distribution to compare to CeRu$_2$Si$_2$.\\

At the threshold of this finite simulation, 24.3\% of the moments are in isolated clusters, and 3\% are in the lattice spanning cluster that is about to fracture. This 0.243 fraction of the 64 million moments are divided over 2.6 million clusters, or 0.041 clusters per lattice site using percolation notation.\cite{stauffer} Removing clusters with a net moment of zero (since they do not show up in magnetization and susceptibility experiments) and clusters of size one (since these are simply isolated moments that are not in an environment that protects them from shielding), we find that the total fraction (compared to the number of sites in the lattice) of uncompensated moments is 0.0147. Thus, the average moment per Ce-ion is of the order of 0.02 $\mu_B$ using the values determined from the uniform susceptibility. This value is in agreement with $\mu$SR experiments.\cite{msr} The average number of uncompensated moments per such cluster equals 1.87. Note, this is the average number of uncompensated moments of all clusters whose net moment does not equal zero and who have more than one cluster member, divided by the total number of such clusters. This large average value is what determines the peak position in the ac-susceptibility: this position is independent of any normalization we perform to determine the value of the average moment from susceptibility of magnetization data, and it is not affected by whether the clusters are fleeting or static in nature. Note that it is actually the full distribution of cluster moments that determines the exact peak position as the averaging involves $\mu_{\textrm{cluster}}^2$. For completeness, we mention that the largest net moment of a cluster in this simulation was a cluster with 94,000 members and a net moment of 326 uncompensated moments (see  Fig. \ref{clusdist}.)\\

The distribution of static clusters at the percolation threshold cannot fully account for the reported\cite{takahashi} discrepancy between the reported moment deduced from the high temperature susceptibility and the saturation magnetization, but it reduces the discrepancy to within a factor of 10. Calculating the average moment, and the average of the square of the moments, following the procedure that we use all lattice sites as our reference as was done in ref [\onlinecite{takahashi}] (rather than using the number of uncompensated moments for normalization), we find a mismatch factor of 18 between the two. Including the two additional factors discussed previously (a factor of 2 to account for the dependence of the saturation level on applied field, and the factor of $\sqrt{3}$), we find an ensuing mismatch about 10 times smaller than the mismatch reported by Takahashi {\it et al.}. Thus, a static cluster distribution successfully explains the overall smallness of the moments when compared to the number of lattice sites, as well as the very large average moment that is responsible for the peak in the ac-susceptibility. It also greatly reduces the unexplained mismatch of a factor of $\sim$1,000 reported by Takahashi {\it et al.} to a mismatch of about 10, but it does not appear to be able to fully get rid of it, even when allowing for the fact that CeRu$_2$Si$_2$ is not exactly at the QCP and that the applied field made an angle of a few tens of degrees with the easy axis. However, we argue next that this remaining mismatch might well be a result of having a system with fleeting clusters rather than static ones.\\

\begin{figure}[t]
    \begin{center}
    \includegraphics*[viewport =140 100 580 450,width =85mm,clip]{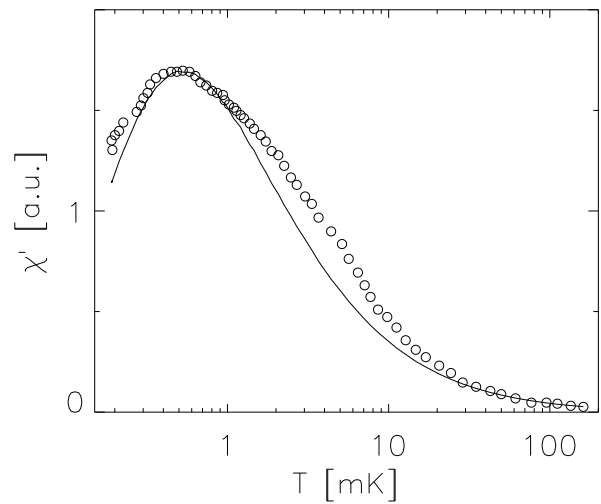}
    \end{center}
    \caption{The measured ac-susceptibility\cite{takahashi} in an applied field of 0.2 mT (circles). The solid line is the result of a computer simulation where each cluster at the percolation threshold was analyzed for uncompensated moments to arrive at the net moment (superspin). The temperature dependence of each superspin was modeled using Eq. \ref{ac}, with each uncompensated moment in these clusters taken to be 1.21 $\mu_B$. This number for the unshielded Ce-moment was inferred from the ratio of the uniform susceptibility and specific heat at low temperatures (see text). An overall scale factor was applied to accommodate the arbitrary units the data were reported\cite{takahashi} in. Note the agreement for both the peak level as well as the higher temperatures, and in addition, the position of the peak is also in agreement between data and simulation. There is no indication of a mismatch of a factor of 100,000.} 
    \label{accluster}
\end{figure}

We now discuss whether the remaining factor of 10 in the mismatch might be (partly) caused by the distinction between a static collection of clusters versus a dynamic one. Takahashi {\it et al.} reported\cite{takahashi} both the dc-susceptibility (in SI units) and the ac-susceptibility (in arbitrary units). Close inspection of Fig. 1 in [\onlinecite{takahashi}] shows that at the lowest fields the agreement with the Curie-Weiss law extends down to lower temperatures (2 mK) for the dc-susceptibility than for the ac-susceptibility (8 mK). This hints at a fleeting distribution of clusters creating a different signal when different measurement techniques are being used: in dc-measurements the sample is physically moved during the measurement cycle, whereas in ac-measurements the signal is being detected instantaneously. In order to investigate whether the dynamics of the cluster distribution might play a role, we scanned in the ac-susceptibility data for a field of 0.20 mT and calculated the ac-response for our collection of static clusters using Eq. \ref{ac}. In order to do so, we assumed an unshielded Ce-moment for the uncompensated Ce-ions of 1.21 $\mu_B$, the number inferred from applying Eq. \ref{lowt}. We applied an overall scale factor to accommodate the arbitrary units the data were reported in. We show the results in Fig. \ref{accluster}. \\

The agreement between the calculated curve based on the simulated cluster distribution and the measured ac-susceptibiity data is highly encouraging. Whereas there clearly is not perfect agreement, there is no trace of a large discrepancy between the higher temperatures (the Curie law region) and the low temperatures (the saturation magnetization region). For reference, the peak position corresponds to a temperature where the magnetization in the static measurements\cite{takahashi} had already reached more than 90\% of its saturation value. Thus, with one scale factor we are able to capture all three moment values as the peak position of the ac-susceptibility does not depend on the overall scale factor. As such, we conclude that the discrepancy of a factor of $\sim$ 100,000 between the highest and lowest moment values as determined from the peak in the ac-susceptibility and the saturation magnetization can be accounted for when taking into account that clusters form, and that these clusters have a fleeting nature.\\

We end this section with a discussion on the in-phase and out-of-phase components of the ac-susceptibility, also referred to as the real part $\chi'$ and the imaginary part $\chi''$ of the complex susceptibility $\chi$. The real part represents the response to the applied external field, whereas the imaginary part yields information about the dissipation of this response. Typically, the imaginary part is small compared to the real part unless the system is close to a phase transition. Takahashi {\it et al.} also showed\cite{takahashi} the data for the imaginary part that displayed an identical temperature dependence to the real part (Fig. 1 in reference [\onlinecite{takahashi}]). In fact, in an earlier publication\cite{earlier} it was reported that $\chi' \approx \chi''$ at a frequency of 16 Hz. Using more measurement frequencies and assuming the validity of the thermodynamic theory for a magnetic system relaxing through coupling with the lattice\cite{casimir}, the authors inferred a relaxation time of approximately 11 ms.\\

The dynamic nature of the Kondo distribution allows for an alternative explanation in CeRu$_2$Si$_2$. Because of the ever-changing distribution, resulting in an ever-changing cluster morphology, we could also view the measured ac-response as clusters appearing spontaneously and aligning with the external field to a degree dictated by the susceptibility. But these clusters also disappear spontaneously, with the result that the signal we observe would not be very strongly dependent of the field-amplitude of the ac-signal, but much more on the static field. If this were the case, then we actually would expect that $\chi' =\chi''$ as clusters appear and disappear, neither in-phase nor out-of-phase with the ac-signal, but with an overall magnetization only depending on the static field. We are not sure what will prove the better explanation for the observed equality of $\chi'$ and $\chi''$ for all fields and temperatures reported in reference [\onlinecite{takahashi}], but if it turns out to be the fleeting nature of clusters in stoichiometric systems, then the ac-susceptibility could act as a litmus test as to their presence.

\section{discussion}

In this paper, we have argued that magnetic clusters form spontaneously in stoichiometric CeRu$_2$Si$_2$ upon cooling because of a distribution of Kondo temperatures. This distribution is the consequence of the exponential sensitivity of the Kondo temperature to interatomic separation. The zero-point-motion of the ions around their equilibrium positions provides a sufficient change in interatomic separations that it leads to a significant variation in Kondo temperature. We have reviewed the abundance of highly-accurate literature data on this system, and through comparison with heavily-doped  Ce(Ru$_{0.755}$Fe$_{0.245}$)$_2$Ge$_2$, we have shown that both systems display the tell-tale signs of magnetic clusters.\\

The presence of magnetic clusters offers a natural explanation for some puzzling observations in CeRu$_2$Si$_2$, some of which were hiding in plain sight. Foremost, there is the discrepancy in the size of the surviving Ce-moment as this was strongly dependent on the method used to ascertain its size. In fact, when we also include the moment that can be inferred from the peak in the ac-susceptibility, then the discrepancy is of the order of 100,000. The presence of clusters offers not merely a qualitative way out of this, but with the aid of computer simulations, we showed that clusters also offer a quantitative solution. We also discussed that cluster formation provides a natural link between the uniform susceptibility and the specific heat, and it might even account for the strange feature observed in stoichiometric CeRu$_2$Si$_2$ that the real and imaginary parts of the susceptibility are identical (at least at the frequency of 16 Hz reported in reference [\onlinecite{takahashi}]). Perhaps most convincingly, the formation of clusters upon lowering the temperature provides an explanation for the observation of identical magnetic correlation lengths along non-identical crystallographic directions when short-range order starts to appear. As far as we are aware, there is not a theory that would predict this to occur.\\

We can also turn our reasoning around: we see no path along which a stoichiometric system can {\it not} harbor magnetic clusters at low temperatures. Zero point motion must result in a distribution of Kondo temperatures at any instant in time, and groups of moments isolated from the rest of the lattice must line up with their neighbors according the the most basic foundations of quantum mechanics that have given us quantized energy levels in confined systems. As such, it should not be a question of whether clusters form, but of whether they have a measurable influence on the response of the system. From our discussion on  Ce(Ru$_{0.755}$Fe$_{0.245}$)$_2$Ge$_2$, Ce(Ru$_{0.97}$Rh$_{0.03}$)$_2$Si$_2$, and CeRu$_2$Si$_2$, clusters not only influence the response, but they dominate it.\\ 

An interesting aspect of clusters in an Ising system is that they still have the degree of freedom to flip. However, Hoyos and Vojta showed\cite{vojta} that the largest clusters may end up being frozen in. Both frozen and reorienting clusters were observed in  Ce(Ru$_{0.755}$Fe$_{0.245}$)$_2$Ge$_2$. Whether they are frozen in, implying that there is an energy barrier to spontaneous reorientations, or whether they are free to flip, this reorientation represents a low lying (in energy) degree of freedom. When the energy required to reorient a cluster is very small compared to the thermal energy, then we can expect to see the equivalent of high-energy physics as the probing energies and thermal energies in a typical experiment greatly exceed the cost of the lowest energy excitations. In experiments, this would be visible as $E/T$ scaling\cite{fridge,schimmel,heitmannprb}. Given this reorientation possibility, we could interpret the field of $H$  = 0.02 mT identified in Fig. \ref{saturation} as the energy barrier. Taking the average moment of isolated clusters to be around 3 $\mu_B$, this barrier would correspond to a thermal energy of about 0.05 K. We are not aware of reports on  CeRu$_2$Si$_2$ that $E/T$ scaling has been observed, although Fig. 2 and 3 in Takahashi {\it et al.} do show indications\cite{takahashi} of $H/T$ scaling.\\

We end with a discussion of Eq. \ref{lowt}. A temperature independent ration of the susceptibility and the linear coefficient of the specific heat is anything but a new finding: the Wilson ratio $R_W$ captures exactly this:
\begin{equation}
R_W \equiv \frac{\pi^2k_B^2\chi}{3\mu_B^2\gamma}\quad \left( =\frac{g_J^2 J_{\textrm{eff}}^2\pi^2}{3\textrm{ln}2} =4.75 g_J^2 J_{\textrm{eff}}^2 \right) ,
\end{equation}
where for the part in brackets we have used Eq. \ref{lowt} to evaluate the ratio. Using the measured values for CeRu$_2$Si$_2$ ($\gamma$ = 0.38 J/mol/K$^2$ and $\chi$($T$ =1.8 K)  = 0.064 $\mu_B$/T/Ce-ion), we find $R_W$  = 6.9 and a corresponding moment of 1.21 $\mu_B$. However, Eq. \ref{lowt} does {\it not} represent the standard Wilson ratio: the Wilson ratio pertains\cite{philip} to the electronic susceptibility and the electronic specific heat. We arrived at a $T$-independent ratio based on the susceptibility associated with localized, unshielded moments and the specific heat reflecting the disappearance of these moments; the two entities that go into our ratio do not involve the effective electron mass nor the density of states at the Fermi-level.\\

We understand that the findings reported in this paper will be met with justified skepticism, mostly because they would necessitate major adjustments to theories valid near a quantum critical point as these theories are based on the response in stoichiometric samples being the same throughout the system. While it is unlikely that this paper ends up being the final word on clusters in stoichiometric systems, the conclusion that restricting ourselves to our current level of theoretical  understanding leaves us with internally inconsistent results on CeRu$_2$Si$_2$ is inescapable.  On a happier note, the findings reported here would extend the validity of disorder-based models to stoichiometric systems, provided that cluster formation is taken into account.

\end{document}